
\baselineskip=18pt
\leftskip 36pt

\noindent
{\bf IS IT POSSIBLE TO EXTEND THE DEFORMED WEYL ALGEBRA W$_q$(n)
TO A HOPF ALGEBRA?}
\vskip 32pt
\noindent

\vskip 32pt
\noindent
T. D. Palev\footnote*{Permanent address: Institute for Nuclear Research
and Nuclear Energy, 1784 Sofia, Bulgaria; E-mail
palev@bgearn.bitnet}

\noindent
International Centre for Theoretical Physics, 34100 Trieste, Italy

\vfill \eject
\vskip 32pt
{\bf Abstract.} We argue that the algebra $W_q(n)$, generated
by $n$ pairs of deformed $q$-bosons, does not allow a Hopf
algebra structure. To this end we show that it is impossible to
define a comultiplication even for the usual, nondeformed case.
We indicate how the comultiplication on $U_q[osp(1/2n)]$ can
be used in order to construct representations of deformed
(not necessarily Hopf) algebras in tensor products of Fock spaces.

\vfill \eject
\vskip 32pt

In his talk in the Workshop on Harmonic Oscillators [1]
L. C. Biedenharn noted that the Weyl algebra $W_q(n)$,
i.e., the infinite-dimensional unital associative algebra,
generated by $n$ pairs of $q-$bosons [2-4], does not allow a complete
Hopf algebra structure and gave a reference in this respect
to me ([1], p.73). This is indeed my point of view, which we discussed with
him. I have, however, no proof of the statement. There are only indications
of the impossibility to define a (nontrivial) comultiplication $\Delta$ on
$W_q(n)$ in a sense of a bialgebra, which I would like to outline in the
present letter.

To begin with we recall that for a bialgebra $U$ the comultiplication
$\Delta$ is an (associative) algebra morphism
$U \rightarrow U \otimes U$, which is also coassociative,
$(id \otimes \Delta)\circ \Delta = (\Delta \otimes id)\circ \Delta$.
Given two representations $\rho_1 :U \rightarrow End(V_1)$
and $\rho_2 :U \rightarrow End(V_2)$ of $U$ in the category of
associative algebras, one can construct, using $\Delta$, a new representation
$\rho$ of $U$ in  the following way. Let $a \in U$ and
$\Delta(a)=\sum_i a_i \otimes b_i $. Then
$\rho(a)=\sum_i \rho_1(a_i) \otimes \rho_2(b_i) \in End(V_1)
\otimes End(V_2)$
gives a representation of $U$ in $V_1 \otimes V_2$. We refer to
this property of $\Delta$ as the main (representation) property of the
comultiplication.

If $L$ is a Lie algebra and $U(L)$ - its universal enveloping algebra,
then $U(L)$ admits only one (nontrivial) comultiplication:

$$\Delta(a)=a\otimes {\bf 1} + {\bf 1}\otimes a \quad \forall a \in L.
\eqno(1)$$

$$\Delta({\bf 1})={\bf 1} \otimes {\bf 1}, \eqno(2)$$

\noindent
where {\bf 1} is the unity of $U(L)$. Indeed, the most general
linear map $\Delta: L \rightarrow U(L) \otimes U(L)$ is

$$\Delta (a)=\sum_i \varphi_i(a) \otimes x_i +
\sum_j y_j \otimes \psi_j(a), \quad \forall a \in L, \; x_i, y_j \in U(L),$$

\noindent
where $\varphi_i, \psi_j:L \rightarrow U(L)$ are linear maps. From
the requirement that $\Delta$ is well defined as a map, i.e.,

$$[\Delta(a), \Delta(b)]=\Delta([a,b]) \quad  \forall a,b \in L,
\eqno(3)$$

\noindent
and using the Poincar\'e-Birkhoff-Witt theorem one derives that
$x_i=y_j={\bf 1}$. Hence
$\Delta(a)=\varphi(a)\otimes {\bf 1} + {\bf 1}\otimes \psi(a)$.
Moreover from (3) it follows also that $\varphi$ and $\psi$ are
endomorphisms of the Lie algebra $L$. The coassociativity yields
$\varphi[\varphi(a)]=\varphi(a)$, $\psi[\psi(a)]=\psi(a)$ for all
$a \in L$. Hence $\varphi=id$ or $0$, $\psi=id$ or $0$.
The comultiplications $\Delta(a)=a \otimes {\bf 1}$,
$\Delta(a)={\bf 1} \otimes a$ and $\Delta(a)=0$ are trivial from
a point of view of the main representation property of the comultiplication.
Therefore the only nontrivial comultiplication is (1).

The first indication  that there exists no comultiplication on $W_q(n)$
is based on the observation that it is impossible to define it
in the form (1) even for the undeformed Weyl
algebra $W(n)$. In order to show this, denote by $H_n$ the
(abstract) Heisenberg Lie algebra, namely the algebra with $2n+1$
generators $c,\; b_1^\pm,\ldots,b_n^\pm$, which satisfy the
commutation relations

$$[b_i^-,b_j^+]=\delta _{ij}c, \quad
[b_i^+,b_j^+]=[b_i^-,b_j^-]=[c,b_j^+]=[c,b_j^-]=0,
\quad i,j=1,\ldots,n. \eqno(4)   $$

\noindent
Let $U(H_n)$ be its universal enveloping algebra (with unity
{\bf 1}). Then eqs.$(1,2)$ define a comultiplication on $U(H_n)$
and in particular

$$\Delta(c)=c\otimes {\bf 1} + {\bf 1}\otimes c. \eqno(5) $$

\noindent
The Weyl algebra $W(n)$ is a factor algebra of $U(H_n)$ with
respect to the ideal generated from the element $c-{\bf 1}$.
In other words, $W(n)$ is the associative algebra with free generators
$c,\; b_1^\pm,\ldots,b_n^\pm$, the relations (4), and one more
relation,

              $$c={\bf 1}. \eqno(6)$$

\noindent
In view of (6) one can write $W(n)$ in its usual form:

$$[b_i^-,b_j^+]=\delta _{ij}{\bf 1}, \quad
[b_i^+,b_j^+]=[b_i^-,b_j^-]=0, \quad i,j=1,\ldots,n. \eqno(7)   $$

\noindent
Now it is evident that the comultiplication (1,2) is
incompatible with the extra relation (6). Indeed, from  (6) one has
$\Delta(c)=\Delta({\bf 1})$, i.e.,
$c\otimes {\bf 1} + {\bf 1}\otimes c={\bf 1} \otimes {\bf 1}$ and
hence $2({\bf 1} \otimes {\bf 1)}={\bf 1} \otimes {\bf 1}$, which is
impossible. Certainly, we would have arrived to the
same conclusion working directly with $W(n)$, i.e., avoiding the
Heisenberg algebra $H_n$.

The second indication of a lack of a comultiplication is based on
the observation that the undeformed Weyl algebra
$W(n)$ has only one (nontrivial irreducible) unitary representation
$\rho_{Fock}$, namely the one, realized in the well known Fock module
$V_{Fock}$. By unitary representation we understand, as usually, that
the representation of the corresponding Weyl group is unitary
(see, for instance [5]).
Assume that $\Delta$ exists on $W(n)$. Then from the main
property of the comultiplication it follows that
$V_{Fock} \otimes V_{Fock}$ is also a $W(n)$ module. On the other
hand, the tensor product of unitary modules is also a unitary module.
Hence $V_{Fock} \otimes V_{Fock}$ has also to be an unitary
module, i.e. to carry a unitary representation of $W(n)$. Therefore
$V_{Fock} \otimes V_{Fock}$ could be only a direct sum of different
copies of Fock modules and a module $V_0$, which is a direct sum of
one-dimensional trivial modules,

$$V_{Fock} \otimes V_{Fock}=V_0 \oplus
\sum_i \oplus (V_{Fock})_i. \eqno(8) $$

In order to show that the decomposition (8) is impossible, we use
the circumstance that $W(n)$ can be viewed as a factor algebra of
another algebraic structure, namely of the universal
enveloping algebra $U[osp(1/2n)]$  of the orthosymplectic Lie
superalgebra $osp(1/2n)$ [6]. The algebra $U[osp(1/2n)]$
is a free algebra of $2n$ generators $B_1^\pm,\ldots,B_n^\pm$,
which satisfy the relations
($\xi, \eta, \epsilon = \pm$ or $\pm 1$, $i,j,k=1,2,\ldots ,n$ ;
$ [x,y]=xy-yx,\; \{x,y\}=xy+yx$)

$$[\{B_i^\xi,B_j^\eta \},B_k^\epsilon]
=(\epsilon - \xi)\delta_{ik}
B_j^\eta + (\epsilon - \eta) \delta_{jk}B_i^\xi.
\eqno(9)$$

\noindent
The operators $B_1^\pm,\ldots,B_n^\pm$ are called para-Bose (pB)
operators [7] and they are  odd generators of $U[osp(1/2n)]$.
The Bose operators satisfy the relations (9);  a  morphism of
$U[osp(1/2n)]$ onto $W(n)$ is given simply by a replacement
$B_i^\pm \rightarrow b_i^\pm$. Therefore any $W(n)$ module and in
particular $V_{Fock}$, $V_{Fock} \otimes V_{Fock}$ and $V_0$ should
be also a $U[osp(1/2n)]$ module. Now it is not difficult to run
into a contradiction with the decomposition (8) and hence with the
assumption that a comultiplication on $W(n)$ exists. We do not go
into the details of the representation theory of the para-Bose operators.
We mention only that $V_{Fock} \otimes V_{Fock}$ is well defined as
a $U[osp(1/2n)]$ module through the (graded) comultiplication

$$\Delta(B_i^\pm)=B_i^\pm \otimes {\bf 1} + {\bf 1}\otimes B_i^\pm
\quad \forall \;  i=1,\ldots,n. \eqno(10)$$

\noindent
The decomposition of $V_{Fock} \otimes V_{Fock}$ into irreducible
$U[osp(1/2n)]$ modules is known [8] and it contains
representations corresponding to the statistics  order of 2.
Such representations are  not present in the right hand
side of the decomposition (8).

We summarize. We have shown that it is impossible to define a
comultiplication (and hence to introduce a Hopf algebra structure)
nether on $W(n)$, considered as an associative algebra, nor in
$W(n)$, considered as an associative superalgebra . It is difficult to
imagine that at $q \not= 1$ there might exists $\Delta$, which
is undefined for $q=1$. Certainly, the arguments we have given above
are only indications, but not a proof of the conjecture that
the deformed Weyl algebra $W_q(n)$ cannot be turned into a
Hopf (super)algebra.

At present it is known how to deform $U[osp(1/2n)]$ to a Hopf
algebra $U_q[osp(1/2n)]$ [9,10]. It has been shown that the latter
is freely generated by deformed pB operators
$B_1^\pm,\ldots,B_n^\pm$ [11,12]. Moreover, it has been
proved that $W_q(n)$ is a factor algebra of $U_q[osp(1/2n)]$ [13],
i.e., that there exists a morphism

$$\pi:\; U_q[osp(1/2n)] \rightarrow W_q(n). \eqno(11) $$

\noindent
If $W_q(n)$ has only one (nontrivial irreducible) representation,
namely the (deformed) Fock representation, then there will be no
difficulty to prove the conjecture. To our knowledge, however, a
proof of the uniqueness of the Fock representation does not exists.

In conclusion we mention that the comultiplication $\Delta$ of
$U_q[osp(1/2n)]$ together with the morphism $\pi$ (see(11))
can replace in several respects the lack of a
comultiplication on $W_q(n)$ as far as representations of deformed
algebras (not necessarily Hopf algebras) are concerned. We have in
mind all subalgebras of $U_q[osp(1/2n)]$, among them $U_q[gl(m)]$,
$m=1,\ldots,n$ (which are Hopf subalgebras), $U_q[sp(2m)]$
$m=1,\ldots,n$ (which are not Hopf subalgebras of $U_q[osp(1/2n)]$;
see, for example, [14] for the case $n=1$)
and several others. In order to be more concrete, define by induction
a morphism

$$\Delta ^{(k)}=[(\otimes_{i=1}^{k-2}id)\otimes \Delta]\circ
\Delta^{(k-1)}, \hskip 12pt \Delta^{(2)}=\Delta, \hskip 12pt
\Delta^{1}=id. \eqno(12) $$

\noindent
of $U_q[osp(1/2n)]$ into $\otimes_{i=1}^{k}U_q[osp(1/2n)]$. The map

$$\pi^{(k)}=(\otimes_{i=1}^{k}\pi) \circ \Delta ^{(k)}:
U_q[osp(1/2n)] \rightarrow \otimes_{i=1}^{k}W_q(n) \eqno(13)  $$

\noindent
is a morphism of $U_q[osp(1/2n)]$ into $\otimes_{i=1}^{k}W_q(n)$
and therefore it is a good substitute for a
comultiplication in a sense that $\pi^{(k)}$ helps  to
construct representations in tensor products of Fock spaces. Indeed
if $A$ is any subalgebra of $U_q[osp(1/2n)]$ and $\rho_{Fock}$ -
the representation of the Weyl algebra $W_q(n)$ in the deformed
Fock space $V_{Fock}$, then the map

$$\rho^{(k)}=(\otimes_{i=1}^{k}\rho_{Fock}) \circ \pi^{(k)} \eqno(14) $$

\noindent
defines a representation of $A$ in $\otimes_{i=1}^k V_{Fock}$. In
this way one can use the Fock representations of $A$  in order to
construct new representations of the same algebra. Thus, despite of the
lack of a comultiplication on $W_q(n)$, one can construct representations
of various deformed algebras in any tensor power of Fock spaces.

\vskip 12pt
I am thankful to Prof. Abdus Salam, the International Atomic Energy
Agency and UNESCO for the kind hospitality at the International
Center for Theoretical Physics.

The research was supported through contract $\Phi$-215 of the
Committee of Science of Bulgaria.
\vfill
\eject

\noindent
{\bf References}

\vskip 12pt
\leftskip 36pt {
\settabs \+  [111111111] & I. Patera, T. D. Palev, Theoretical
   interpretation of the experiments on the elastic \cr

\+ \hskip 36 pt [1] & Biedenharn L C in 1993 {\it Workshop on Harmonic
        Oscillators} 69 \cr
\+    & (NASA Conference Publications 3197,
        Edited by Han D, Kim Y S and Zachary W W)  \cr

\+ \hskip 36 pt [2] & Biedenharn  L C  1989 {\it J. Phys. A} {\bf 22}
         L873 \cr

\noindent
\+ \hskip 36 pt [3] & Macfarlane  A J  1989 {\it J. Phys. A} {\bf 22}
        4581 \cr

\noindent
\+ \hskip 36 pt [4] & Sun  C P  and Fu  H C  1989
        {\it J. Phys. A} {\bf 22} L983 \cr

\+ \hskip 36 pt  [5] & Folland G B 1989 {\it Harmonic Analysis in
        Phase Space} \cr
\+ \hskip 36 pt   & (Princeton Univ. Press, Princeton, New Jersey) \cr

\+ \hskip 36 pt [6] & Ganchev A and Palev T D 1980 {\it J. Math. Phys.}
      {\bf 21} 797 \cr

\+ \hskip 36 pt  [7] & Green H S 1953 {\it Phys. Rev.} {\bf 90}  270 \cr

\+ \hskip 36 pt  [8] & Ohnuki Y and Kamefuchi 1982 {\it Quantum Field
        Theory and Parastatistics} \cr
\+ \hskip 36 pt   & (Univ. of Tokyo Press, Springer-Verlag, Berlin) \cr

\+ \hskip 36 pt  [9] & Tolstoy  V N  1990 {\it Lect. Notes Phys.}
          {\bf 370} 118 \cr

\+ \hskip 36 pt  [10] & Floreanini R, Spiridonov V P  and Vinet L
        1991 {\it Comm. Math. Phys.} {\bf 137} 149 \cr

\+ \hskip 36 pt [11] & Hadjiivanov L K 1993 Quantum deformations of Bose
        parastatistics {\it Preprint} ESI 20, Vienna \cr

\+ \hskip 36 pt [12] & Palev T D 1993 Quantisation of $U_q[osp(1/2n)]$
          with deformed para-Bose operators,  \cr
\+ \hskip 36 pt & {\it Preprint} University of Ghent TWI-93-24.\cr

\+ \hskip 36 pt  [13] & Palev T D 1993 A superalgebra morphism of
        $U_q[osp(1/2n)]$  onto the deformed oscillator \cr
\+ \hskip 36pt   & superalgebra $W_q(n)$, {\it Preprint}
        University of Ghent TWI-93-17,
        {\it Lett. Math. Phys.} (to appear) \cr

\+ \hskip 36 pt  [14] & Kulish P P and Reshetikhin N Yu 1989
        {\it Lett. Math. Phys.} {\bf 18} 143 \cr

\end